\documentclass{article}

\usepackage{arxiv}

\usepackage[utf8]{inputenc} 
\usepackage[T1]{fontenc}    
\usepackage{hyperref}       
\usepackage{url}            
\usepackage{booktabs}       
\usepackage{amsfonts}       
\usepackage{nicefrac}       
\usepackage{microtype}      
\usepackage{lipsum}		
\usepackage{graphicx}
\usepackage{natbib}
\usepackage{doi}

\usepackage{amsmath,amsthm,amssymb}
\usepackage{xifthen}%
\usepackage{xparse}
\usepackage{tikz}

\title{Strategyproof and Proportional Chore Division for Piecewise Uniform Preferences}

\author{ {David Francis} \\
	Department of Computer Science\\
	University of Toronto\\
	\texttt{delta.david.john.francis@gmail.com} \\
}

\renewcommand{\shorttitle}{Strategyproof and Proportional Chore Division}

\hypersetup{
pdftitle={Strategyproof and Proportional Chore Division for Piecewise Uniform Preferences},
pdfsubject={cs.GT},
pdfauthor={David Francis},
pdfkeywords={Cake cutting, Strategy-proofness, Chore division, Mechanism design, Fair division},
}

\NewDocumentCommand{\INTERVALINNARDS}{ m m }{
    #1 {,} #2
}
\NewDocumentCommand{\interval}{ s m >{\SplitArgument{1}{,}}m m o }{
    \IfBooleanTF{#1}{
        \left#2 \INTERVALINNARDS #3 \right#4
    }{
        \IfValueTF{#5}{
            #5{#2} \INTERVALINNARDS #3 #5{#4}
        }{
            #2 \INTERVALINNARDS #3 #4
        }
    }
}

\newtheorem{thm}{Theorem}
\newtheorem{lem}[thm]{Lemma}

\begin{document}
\maketitle

\begin{abstract}
Chore division is the problem of fairly dividing some divisible, undesirable bad, such as a set of chores, among a number of players. Each player has their own valuation of the chores, and must be satisfied they did not receive more than their fair share. In this paper, I consider the problem of strategyproof chore division, in which the algorithm must ensure that each player cannot benefit from mis-representing their position. I present an algorithm that performs proportional and strategyproof chore division for any number of players given piecewise uniform valuation functions.
\end{abstract}

\section{Introduction}

Cake cutting is the problem of dividing a heterogeneous, divisible good such as a cake among some number of players such that each player is satisfied with their piece. A valid algorithm must ensure that each player must value their piece at at least $\frac{1}{n}$ the value of the whole cake (the proportionality condition). The two-player solution ``I cut, you choose'' is ancient. Solutions for any number of players date back to at least 1949 (\cite{stein1949}).

Chore division is the inverse problem to cake cutting, in which the object to be divided is undesirable and each player must receive at most their fair shore (confusingly, the term chore division is also used to refer to the division of indivisible bads, the inverse of the estate division problem). Chore division might model the allocation of chores within a household, work shifts at a hospital, or liabilities in a bankrupt company. Chore division and cake cutting are surprisingly distinct problems; algorithms from one cannot easily be applied to the other. For most questions in cake cutting there exist parallel questions in chore division; for instance, an $n$-person envy-free cake cutting algorithm, in which each player is satisfied that no other player has received a bigger piece in their estimation, was found by \cite{brams1995envy}, while the equivalent chore-division result was found by \cite{peterson2009n}. There have also been some work in the modeling of mixed manna or burnt cakes: goods which may have positive or negative value, such as the existence results of \cite{bogomolnaia2017competitive} and \cite{segal2017fairly}.

Strategyproofness is a concept in mechanism design for many problems of group decision making. A strategyproof algorithm (also known as a truthful or incentive compatible algorithm) is one in which no player can improve their position by lying about their preferences. For example, a strategyproof voting algorithm would not allow voters to change the result of the election to a more preferred candidate by submitting a ballot containing anything other than their true ranking of the candidates.

To apply strategyproofness to cake cutting and chore division algorithms, some restrictions must be placed on the players' valuations. The standard model of cake cutting is to treat the interval $\interval[{0,1}]$ as the cake and have the pieces of cake allocated to each player be subsets of that interval. Under this model, a players' preferences can be represented by a valuation function, which assigns a marginal value (or cost, in the case of chore division) to each real number in the cake interval. 

Two common restricted classes of valuation function are piecewise constant and piecewise uniform. A piecewise constant valuation function divides the cake into some number of intervals, with the marginal value constant in each interval. A piecewise uniform valuation function is the same, with the additional restriction that the marginal values in each interval must either be $0$ or $1$. In other words, a piecewise uniform valuation function divides the cake into two sections: one with uniformly distributed value and one with zero value. These kinds of restricted valuation functions are necessary in the discussion of strategyproof algorithms; \cite{mossel2010truthful} proved that for arbitrary valuation functions, no deterministic strategyproof cake cutting algorithm exists.

\cite{maya2012incentive} characterized the problem of strategyproof cake cutting with piecewise uniform valuation functions, finding a bound on the social welfare. \cite{chen2013truth} found the first strategyproof algorithm for cake cutting with piecewise uniform valuation functions, which was extended by \cite{aziz2014cake}. The algorithm in \cite{chen2013truth} is also envy-free.

As cake cutting and chore division algorithms have continuous inputs and outputs, a specialized complexity model is needed to measure their complexity. The standard approach is the model of \cite{robertson1998cake}. This treats an algorithm as a series of cut and evaluate queries. A cut query asks a player to divide some piece of cake into two pieces such that the values of the pieces in that player's estimation are in some given ratio. An evaluate query asks a player to give their valuation of some given piece of cake. 

There is an important distinction between bounded algorithms and unbounded algorithms. A bounded algorithm is one for which there is some function $f(n)$ such that running the algorithm with $n$ players will never take more than $f(n)$ queries. The envy-free allocation algorithms found by \cite{brams1995envy} and \cite{peterson2009n} are unbounded. Bounded envy-free algorithms for cake cutting and chore division were found by \cite{aziz2016discrete} and \cite{dehghani2018envy} respectively. The algorithm from \cite{chen2013truth} is unbounded; in fact \cite{kurokawa2013cut} proved that there is no cake cutting algorithm that is strategyproof, bounded, and envy-free.

\subsection{My Results}

I present a deterministic algorithm for chore division with piecewise uniform valuation functions that is strategyproof and proportional. The algorithm allocates all of the chores with no overlap between assignments. It is pareto-efficient, but not envy-free.

\section{Preliminaries}

The chores to be divided will be represented by intervals of the real numbers. A \textit{segment of chores} is the union of a finite number of finitely large disjoint intervals. The length of the segment of chores $C$, written $|C|$, is the sum of the lengths of its component intervals $\sum_{\interval[{x_i,y_i}] \in C} y_i - x_i$. The union of or intersection of two segments of chores is itself a segment of chores. I will use $\bar{C}$ to denote the set of real numbers not in $C$.

There is a segment of chores $C$ that must be divided among $n$ players $\{P_1, P_2, ..., P_n\}$. $C$ must be partitioned into $n$ segments of chores $\{D_1, D_2, ...D_n\}$. 

Each player has their own valuation of the chores; they might consider some chores deeply unpleasant and want to avoid them at all costs, while seeing other chores as relatively painless. In the general case, this can be represented by a valuation function that assigns a real marginal cost to each point in $C$. However, in this paper I will consider only the case of \textit{piecewise uniform} valuations. Each player divides the chores into two segments: a segment to which they assign uniformly distributed cost and a segment to which they assign zero cost.

The segment of chores that the player $P_i$ assigns positive cost to is $P_i$'s valuation set $V_i$, a (possibly empty) subset of $C$. If $P_i$ is assigned the segment $D_i$, the cost to $P_i$ is $|D_i \cap V_i|$ of a possible $|V_i|$. 

The canonical version of the problem has each player trying to receive at most $\frac{1}{n}$ of the chores in their estimation, but it will be convenient to define the problem more generally, with each player having an assigned workload of the chores $w_i$, with the $\{w_1, w_2, ...,w_n\}$ positive and summing to $1$.  

A \textit{chore division algorithm} is a function $f$. The arguments to the function are a segment of chores to be divided $C$, an integer number of players $n$, $n \geq 2$, a valuation set for each player $V = \{V_1, V_2, ...V_n\}$, with each valuation set a segment and a subset of $C$, and an assigned workload for each player $w = \{w_1, w_2, ...,w_n\}$, with each workload a positive real number and the workloads all summing to $1$. The result of the function is a partition of $C$ into $n$ segments $\{D_1, D_2, ...D_n\}$.

An algorithm is \textit{proportional} if each player receives at most $w_i$ of the chores they wanted to avoid. For all $i$, $|V_i \cap D_i| \leq w_i |V_i|$.

Given some result of the algorithm $f(C, n, V, w) = \{D_1, D_2, D_3, ..., D_i, ...\}$, player $P_i$ might attempt to improve their standing by reporting a false valuation set $V_i^*$. If $V^*$ is the list of valuation sets that is the same as $V$ except that $V_i$ is replaced by $V_i^*$, the result of the deception is the new partition $f(C, n, V^*, w) = \{D_1^*, D_2^*, D_3^*, ..., D_i^*, ...\}$. An algorithm $f$ is \textit{strategyproof} if, for any $C$, $n$, $V$, $p$, and $V_i^*$, $|V_i \cap D_i| \leq |V_i \cap D_i^*|$.

\section{The Split Rulership Algorithm}

The algorithm has as its input a segment of chores to be divided $C$, an integer number of players $n$, $n \geq 2$, a valuation set for each player $V = \{V_1, V_2, ...V_n\}$, with each valuation set a segment and a subset of $C$, and an assigned workload for each player $w = \{w_1, w_2, ...,w_n\}$, with each workload a positive real number and the workloads all summing to $1$. The algorithm is recursive and asymmetrical; the ordering of the players is relevant.

The first two players, $P_1$ and $P_2$, split the chores $C$ at a point $k$ into two segments. $P_1$ rules the left segment $\interval({-\infty, k}) \cap C$, and $P_2$ rules the right segment $\interval[{k, \infty}) \cap C$.

$P_1$ is exempted from doing any of the chores they marked as having cost within the segment they rule. In exchange, they are assigned all of the chores within the segment they rule that they did not mark. Similarly, $P_2$ is exempted from $\interval[{k, \infty}) \cap C \cap V_2$ but must take $\interval[{k, \infty}) \cap C \cap \bar{V_2}$.

$k$ is chosen such that $P_1$ and $P_2$ exempt themselves from equal length of chores, weighted by their assigned workload.

\[\text{Let } k \text{ be the lowest real number such that } w_1|\interval({-\infty, k}) \cap V_1| = w_2|\interval[{k, \infty}) \cap V_2|\]

Call the chores $P_1$ has exempted themselves from $C_L$ and the chores $P_2$ has exempted themselves from $C_R$.

\[C_L = \interval({-\infty, k}) \cap V_1\]
\[C_R = \interval[{k, \infty}) \cap V_2\]

$C_L$ must be divided among the $n - 1$ players other than $P_1$, and $C_R$ among the $n - 1$ players other than $P_2$.

If $n = 2$, there is only one player remaining for $C_L$ and for $C_R$. That player is assigned all of that segment.

\[\text{If } n = 2\text{, } D_1 = (\interval({-\infty, k}) \cap \bar{V_1} \cap C) \cup C_R\]
\[\text{If } n = 2\text{, } D_2 = (\interval[{k, \infty}) \cap \bar{V_2} \cap C) \cup C_L\]

Otherwise $P_2$, who ruled the right segment in the division of $C$, will rule the right segment in the division of $C_L$, with the next player in line $P_3$ ruling the left segment. Similarly $P_1$ will rule the left segment in the division of $C_R$, with $P_3$ ruling the right segment.

\[V_L = \{V_3 \cap C_L, V_2 \cap C_L, V_4 \cap C_L, ...\}\]
\[V_R = \{V_1 \cap C_R, V_3 \cap C_R, V_4 \cap C_R, ...\}\]

Since $P_1$ is not participating in the division of $C_L$, someone must take on their assigned workload $w_1$ to keep the workloads adding to $1$. $P_2$ will do so; their assigned workload in $C_L$ will be $w_1 + w_2$. All other players' workloads are unchanged. Symmetrically, in $C_R$ $P_1$ will instead take on $P_2$'s workload.

\[w_L = \{w_3, w_1 + w_2, w_4, ...\}\]
\[w_R = \{w_1 + w_2, w_3, w_4, ...\}\]

The algorithm is then run recursively on $C_L$ and on $C_R$.

\[\{D_3^L, D_2^L, D_4^L, ...\} = f(C_L, n - 1, V_L, w_L)\]
\[\{D_1^R, D_3^R, D_4^R, ...\} = f(C_R, n - 1, V_R, w_R)\]

\[\text{If } n > 2\text{, } D_1 = (\interval({-\infty, k}) \cap \bar{V_1} \cap C) \cup D_1^R\]
\[\text{If } n > 2\text{, } D_2 = (\interval[{k, \infty}) \cap \bar{V_2} \cap C) \cup D_2^L\]
\[\text{If } n > 2\text{, for any } i > 2\text{, } D_i = D_i^L\cup D_i^R\]

\begin{figure}
\centering
\begin{tikzpicture}

\node[anchor = east] at (0,0) {$V_1$};
\node[anchor = east] at (0,-0.5) {$V_2$};
\node[anchor = east] at (0,-1) {$V_3$};

\node[anchor = east] at (0,-2) {$C$};
\node[anchor = east] at (0,-2.5) {$C_L$};
\node[anchor = east] at (0,-3) {$C_R$};

\node[anchor = east] at (0,-4) {$D_1$};
\node[anchor = east] at (0,-4.5) {$D_2$};
\node[anchor = east] at (0,-5) {$D_3$};

\draw[|-|] (0,0) -- (1,0);
\draw[|-|] (2,0) -- (3,0);
\draw[|-|] (5.5,0) -- (8,0);
\draw[|-|] (1.5,-0.5) -- (4,-0.5);
\draw[|-|] (6,-0.5) -- (7,-0.5);
\draw[|-|] (0.5,-1) -- (7,-1);

\draw[|-|] (0,-2) -- (8,-2);
\draw[|-|] (0,-2.5) -- (1,-2.5);
\draw[|-|] (2,-2.5) -- (3,-2.5);
\draw[|-|] (3,-3) -- (4,-3);
\draw[|-|] (6,-3) -- (7,-3);

\draw[thin] (3,-1.9) -- (3,-2.1);
\node[anchor = south] at (3,-2) {$k$};
\draw[thin] (2.5,-2.4) -- (2.5,-2.6);
\node[anchor = south] at (2.5,-2.5) {$k$};
\draw[thin] (6.333,-2.9) -- (6.333,-3.1);
\node[anchor = south] at (6.333,-3) {$k$};

\draw[|-|] (1,-4) -- (2,-4);
\draw[|-|] (3,-4) -- (4,-4);
\draw[|-|] (6.333,-4) -- (7,-4);

\draw[|-|] (0.5,-4.5) -- (1,-4.5);
\draw[|-|] (2,-4.5) -- (2.5,-4.5);
\draw[|-|] (4,-4.5) -- (6,-4.5);
\draw[|-|] (7,-4.5) -- (8,-4.5);

\draw[|-|] (0,-5) -- (0.5,-5);
\draw[|-|] (2.5,-5) -- (3,-5);
\draw[|-|] (6,-5) -- (6.333,-5);

\end{tikzpicture}
\caption{A sample run of the split rulership algorithm. $w_1 = w_2 = w_3 = \frac{1}{3}$}
\label{dia_sample}
\end{figure}
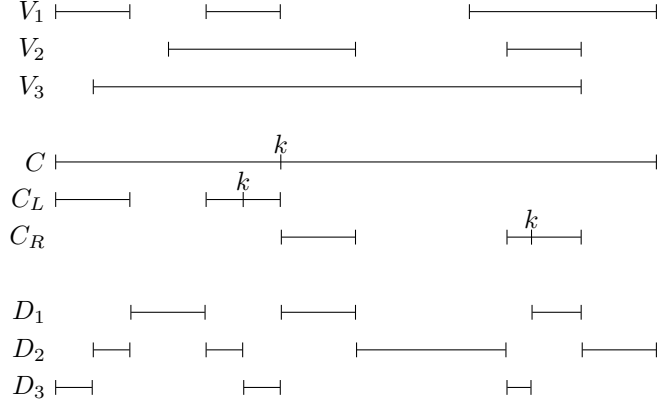

\subsection{Proportionality}

\begin{lem}

The split rulership algorithm is proportional for $n = 2$

\end{lem}

First, consider the case of $P_1$. $D_1$, the chores assigned to $P_1$, consist of the chores to the left of $k$ that $P_1$ has not marked and the the chores to the right of $k$ that $P_2$ has marked.

\[D_1 = (\bar{V_1} \cap \interval({-\infty, k}) \cap C) \cup  (V_2 \cap \interval[{k, \infty}))\]

The cost paid by $P_1$ in their own estimation is $|V_1 \cap D_1|$. As $P_1$ is exempted from all chores they have marked to the left of $k$, $V_1 \cap D_1$ lies entirely to the right of $k$.

\[|V_1 \cap D_1| = |V_1 \cap ((\bar{V_1} \cap \interval({-\infty, k}) \cap C) \cup  (V_2 \cap \interval[{k, \infty})))| \]
\[|V_1 \cap D_1| = |V_1 \cap V_2 \cap \interval[{k, \infty})| \]

These chores are a subset of $C_R$.

\[|V_1 \cap D_1| \leq |V_2 \cap \interval[{k, \infty})| \]

We have that $w_1|V_1 \cap \interval({-\infty, k})| = w_2|V_2 \cap \interval[{k, \infty})|$

\[|V_1 \cap D_1| \leq \frac{w_1}{w_2}|V_1 \cap \interval({-\infty, k})| \]

Adding $\frac{w_1}{w_2}|V_1 \cap D_1|$ to each side,

\[(1 + \frac{w_1}{w_2})|V_1 \cap D_1| \leq \frac{w_1}{w_2}(|V_1 \cap \interval({-\infty, k})| + |V_1 \cap D_1|) \]

Since $V_1 \cap D_1$ is entirely to the right of $k$, $V_1 \cap \interval({-\infty, k})$ and $V_1 \cap D_1$ are non-intersecting subsets of $V_1$.

\[(1 + \frac{w_1}{w_2})|V_1 \cap D_1| \leq \frac{w_1}{w_2}|V_1| \]

Using the fact that, if $n = 2$,  $w_1 + w_2 = 1$.

\[\frac{1}{w_2}|V_1 \cap D_1| \leq \frac{w_1}{w_2}|V_1| \]
\[|V_1 \cap D_1| \leq w_1|V_1| \]

A similar, symmetrical argument holds for $P_2$.

\begin{thm}

The split rulership algorithm is proportional

\end{thm}

Since the $n = 2$ case is already taken care of, we may assume $n > 2$. Beginning with $P_1$, $D_1$ consists of the chores to the left of $k$ that $P_1$ did not mark and $P_1$'s assigned chores from the division of $C_R$, $D_1^R$.'

\[|V_1 \cap D_1| = |V_1 \cap ( (\bar{V_1} \cap \interval({-\infty, k}) ) \cup  D_1^R)|\]

Once again, all of $V_1 \cap D_1$ lies entirely to the right of $k$.

\[|V_1 \cap D_1| = |V_1 \cap D_1^R|\]

The division of $C_R$ was between $n - 1$ players. By induction on the number of players, we may take as given that the algorithm is proportional when dividing among $n - 1$ players. $P_1$'s assigned workload within $C_R$ is $w_1 + w_2$, so $P_1$ must have received at most $w_1 + w_2$ of $V_1 \cap C_R$.

\[|V_1 \cap D_1^R| \leq (w_1 + w_2)|V_1 \cap V_2 \cap \interval[{k, \infty})|\]

$V_1 \cap V_2 \cap \interval[{k, \infty})$ is a subset of $C_R$, and $|C_R| = \frac{w_1}{w_2}|C_L|$

\[|V_1 \cap V_2 \cap \interval[{k, \infty})| \leq |V_2 \cap \interval[{k, \infty})| = \frac{w_1}{w_2}|V_1 \cap \interval({-\infty, k})|\]

Adding $\frac{w_1}{w_2}|V_1 \cap V_2 \cap \interval[{k, \infty})|$ to each side,

\[(1 + \frac{w_1}{w_2})|V_1 \cap V_2 \cap \interval[{k, \infty})| \leq \frac{w_1}{w_2}(|V_1 \cap \interval({-\infty, k})| + |V_1 \cap V_2 \cap \interval[{k, \infty})|)\]

Since $|V_1 \cap \interval({-\infty, k})|$ and $|V_1 \cap V_2 \cap \interval[{k, \infty})|$ are non-intersecting subsets of $V_1$, 

\[(1 + \frac{w_1}{w_2})|V_1 \cap V_2 \cap \interval[{k, -\infty})| \leq \frac{w_1}{w_2}|V_1|\]
\[(w_1 + w_2)|V_1 \cap V_2 \cap \interval[{k, \infty})| \leq w_1|V_1|\]
\[|V_1 \cap D_1| \leq w_1|V_1| \]

Symmetrically, the same argument holds for $P_2$.

All that remains is the case of players other than $P_1$ and $P_2$. For any $P_i$, $i > 2$, $D_i$ consists of $P_i$'s assigned portions from the division of $C_L$ and $C_R$, $D_i^L$ and $D_i^R$.

\[|V_i \cap D_i| = |V_i \cap D_i^L| + |V_i \cap D_i^R|\]

$P_i$'s assigned workload in both of those division is $w_i$. By induction, we may take as given that both of these divisions were proportional, and that $P_i$ received at most $w_i$ of their marked chores in each of those segments. 
\[|V_i \cap D_i| \leq w_i (|V_i \cap V_1 \cap \interval[{-\infty, k})| + |V_i \cap V_2 \cap \interval[{k, \infty})|)\]
Since those two segments are non-intersecting subsets of $V_i$,
\[|V_i \cap D_i| \leq w_i |V_i|\]

\subsection{Strategyproofness}

Consider first the case of $P_1$, the first player. For some input to the split rulership algorithm $(C, n, V, w)$ and some alternate valuation set for $P_1$ $V_1^*$, let $V^*$ be the same as $V$ except $V_1$ is replaced by $V_1^*$, $\{D_1, D_2, ... ,D_n\} = f(C, n, V, w)$, and $\{D_1^*, D_2^*, ..., D_n^*\} = f(C, n, V^*, p)$.

$P_1$ rules the left segment of $C$ and so is exempted from $C_L$, but must participate in the division of $C_R$. If $C_{RL}$ and $C_{RR}$ are the left and right exempted segments in the division of $C_R$, $P_1$ is exempted from $C_{RL}$ and participates in the division of $C_{RR}$. $P_1$ also exempts themself from $C_{RRL}$, $C_{RRRL}$, and so on for all $n-1$ rounds. In the $n-1$th round of division, $P_1$ and $P_n$ will be the only two players remaining, and $P_1$ will be assigned some of the chores $P_n$ rules. 

Let us introduce some more convenient notation. Let $C_1$ be the original $C$, and $k_1$ the $k$ where it is divided. Let $C_2$ be the $C_R$ in the division of $C$, $C_3$ the $C_R$ in the division of $C_2$, and so on. We may ignore the division of the $C_L$ of each of these iterations, as $P_1$ is not assigned any of them.

\[C_1 = C\]
\[C_{j + 1} = C_j \cap V_{j + 1} \cap \interval[{k_j, \infty})\]

Or, non-recursively defined,

\[C_j = C \cap V_2 \cap V_3 \cap ... \cap V_j \cap \interval[{k_{j - 1}, \infty})\]

With $C_1^*, C_2^*, ... C_{n-1}^*$ and $k_1^*, k_2^*, ... k_{n-1}^*$ defined equivalently for the algorithm as run on $V^*$. Note that all the valuation sets except $P_1$'s are unchanged in this case.

\[C_j^* = C \cap V_2 \cap V_3 \cap ... \cap V_j \cap \interval[{k_{j - 1}^*, \infty})\]

\begin{figure}
\centering
\begin{tikzpicture}

\node[anchor = east] at (0,0) {$D_1$};
\node[anchor = east] at (0,1) {$C_{n-1}$};
\node[anchor = east] at (0,2) {$C_3$};
\node[anchor = east] at (0,2.5) {$C_2$};
\node[anchor = east] at (0,3) {$C_1$};
\node[anchor = east] at (0,4) {$V_1$};

\draw[|-|] (1,0) -- (1.5,0);
\draw[|-|] (3,0) -- (3.4,0);
\draw[|-|] (6.5,0) -- (6.75,0);

\draw[|-|] (6,1) -- (6.8,1);

\draw[|-|] (3.5,2) -- (4,2);
\draw[|-|] (5,2) -- (5.7,2);
\draw[|-|] (6,2) -- (7,2);

\draw[|-|] (2,2.5) -- (2.5,2.5);
\draw[|-|] (3,2.5) -- (4,2.5);
\draw[|-|] (5,2.5) -- (7,2.5);

\draw[|-|] (0,3) -- (7,3);

\draw[|-|] (0,4) -- (1,4);
\draw[|-|] (1.5,4) -- (2.7,4);
\draw[|-|] (3.4,4) -- (4.9,4);
\draw[|-|] (5.8,4) -- (7,4);

\draw[thin] (2,2.9) -- (2,3.1);
\node[anchor = south] at (2,3) {$k_1$};
\draw[thin] (3.5,2.4) -- (3.5,2.6);
\node[anchor = south] at (3.5,2.5) {$k_2$};
\draw[thin] (3.8,1.9) -- (3.8,2.1);
\node[anchor = south] at (3.8,2) {$k_3$};
\draw[thin] (6.5,0.9) -- (6.5,1.1);
\node[anchor = south] at (6.5,1) {$k_{n-1}$};

\node[anchor = south] at (5,0) {$...$};
\node[anchor = south] at (6,1.5) {$...$};

\end{tikzpicture}
\caption{The split rulership algorithm as seen by $P_1$}
\label{dia_sproof}
\end{figure}
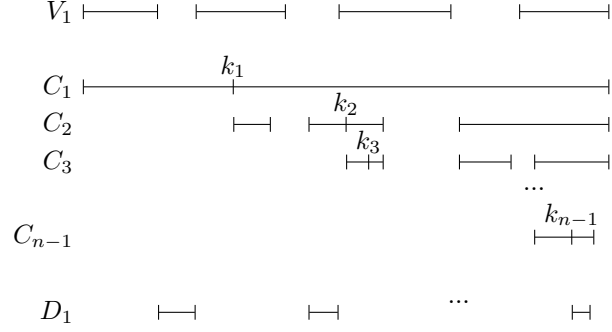

From $C_1$ up to $C_{n-1}$, $P_1$ is assigned the chores to the left of $k_i$ that are outside of $V_1$. In the division of $C_{n-1}$, $P_1$ is also assigned the chores that $P_n$ exempts themselves from to the right of $k_{n-1}$.

\[D_1 = (C_1 \cap \interval({-\infty, k_1}) \cap \bar{V_1}) \cup (C_2 \cap \interval[{k_1, k_2}) \cap \bar{V_1}) \cup ... \cup (C_{n - 1} \cap \interval[{k_{n - 2}, k_{n - 1}}) \cap \bar{V_1}) \cup (C_{n - 1} \cap \interval[{k_{n - 1}, \infty}) \cap V_n)\]

All the chores $P_1$ is assigned that $P_1$ assigns cost to are to the right of $k_{n-1}$.

\[V_1 \cap D_1 = C_{n - 1} \cap \interval[{k_{n - 1}, \infty}) \cap V_n \cap V_1\]

These are precisely the chores that all players marked that lie to the right of $k_{n -1}$.

\[V_1 \cap D_1 = V_1 \cap V_2 \cap ... \cap V_n \cap \interval[{k_{n - 1}, \infty})\]

\begin{lem}

If $k_{n-1}^* < k_{n-1}$, $|V_1 \cap D_1| \leq |V_1 \cap D_1^*|$

\end{lem}

Suppose $k_{n-1}^* < k_{n-1}$. $P_1$ will be assigned all chores to the right of $k_{n-1}^*$ that all players other than $P_1$ marked.

\[D_1^* \supseteq  C_{n - 1}^* \cap \interval[{k_{n - 1}^*, \infty}) \cap V_n\]
\[D_1^* \supseteq V_2 \cap V_3 \cap ... \cap V_n \cap \interval[{k_{n - 1}^*, \infty}) \]
\[|V_1 \cap D_1^*| \geq |V_1 \cap V_2 \cap ... \cap V_n \cap \interval[{k_{n - 1}^*, \infty})|\]

As $k_{n-1}^*$ is to the left of $k_{n-1}$

\[|V_1 \cap D_1^*| \geq |V_1 \cap V_2 \cap ... \cap V_n \cap \interval[{k_{n - 1}, \infty})|\]
\[|V_1 \cap D_1^*| \geq |V_1 \cap D_1|\]

Then, for $P_1$ to get any advantage from their deception, $k_{n-1}^* \geq k_{n-1}$, and from this point on I will assume that this is true. Some of the $k_i^*$ may be less than the equivalent $k_i$, some may be greater, and some may be the same. Let $k_q^*$ be the first of these $k_i^*$ for which all the $k_i^*$ from $k_q^*$ up to $k_{n-1}^*$ are greater than or equal to their respective $k_i$.

\[k_{q - 1}^* < k_{q - 1} \text{ or } q = 1\]
\[\text{For all } j \geq q \text{, } k_j^* \geq k_j \]

If $q = 1$, the argument proceeds slightly differently. In that case, replace $k_{q-1}$ and $k_{q-1}^*$ with $-\infty$ wherever they appear below.

The portion of $D_1$ that lies within $C_q$ is

\[D_1 \cap C_q = (C_q \cap \interval[{k_{q-1}, k_q}) \cap \bar{V_1}) \cup (C_{q+1} \cap \interval[{k_q, k_{q+1}}) \cap \bar{V_1}) \cup ... \cup (C_{n - 1} \cap \interval[{k_{n - 2}, k_{n - 1}}) \cap \bar{V_1}) \cup (C_{n - 1} \cap \interval[{k_{n - 1}, \infty}) \cap V_n)\]
\[D_1 \cap C_q = (\bar{V_1} \cap ((C_q \cap \interval[{k_{q-1}, k_q})) \cup (C_{q+1} \cap \interval[{k_q, k_{q + 1}})) \cup ... \cup (C_{n-1} \cap \interval[{k_{n-2}, k_{n-1}})))\cup(C_{n-1}\cap \interval[{k_{n - 1}, \infty})\cap V_n)\]

Let $C_H$ be defined as follows

\[C_H = (C_q \cap \interval[{k_{q-1}, k_q})) \cup (C_{q+1} \cap \interval[{k_q, k_{q + 1}})) \cup ... \cup (C_{n-1} \cap \interval[{k_{n-2}, k_{n-1}}))\]
\[D_1 \cap C_q = (\bar{V_1} \cap C_H) \cup (C_{n-1}\cap \interval[{k_{n - 1}, \infty})\cap V_n)\]

And similarly,

\[C_H^* = (C_q^* \cap \interval[{k_{q-1}^*, k_q^*})) \cup (C_{q+1}^* \cap \interval[{k_q^*, k_{q + 1}^*})) \cup ... \cup (C_{n-1}^* \cap \interval[{k_{n-2}^*, k_{n-1}^*}))\]
\[D_1^* \cap C_q^* = (\bar{V_1^*} \cap C_H^*) \cup (C_{n-1}^*\cap \interval[{k_{n - 1}^*, \infty})\cap V_n)\]

\begin{lem}

$C_H \subseteq C_H^*$
\label{eq:h_to_h}

\end{lem}

Suppose $h$ is a real number in $C_H$. $k_{q-1} < h < k_{n-1}$, and there is some integer $i$, $q \leq i \leq n-1$, for which $k_{i-1} \leq h < k_i$.

\[h \in C_i \cap \interval[{k_{i-1}, k_i})\]
\[h \in C \cap V_2 \cap V_3 \cap ... \cap V_i\]

$k_{q-1}^* \leq k_{q-1}$ and $k_{n-1}^* > k_{n-1}$, so $k_{q-1}^* < h < k_{n-1}^*$. Then there is also some integer $j$, $q \leq j \leq n-1$, for which $k_{j-1}^* \leq h < k_j^*$. Because all of the $k_j^*$ are to the right of their respective $k_j$, $i \leq j$.

\[h \in C \cap V_2 \cap V_3 \cap ... \cap V_j\]
\[h \in C_j^* \cap \interval[{k_{j-1}^*, k_j^*})\]
\[h \in C_H^*\]

As this is true for any real number in $C_H$,

\[ C_H \subseteq C_H^* \]

\begin{lem}

$C_{n-1} \cap \interval[{k_{n-1}, k_{n-1}^*}) \subseteq C_H^* \cap \interval[{k_{n-1}, k_{n-1}^*})$
\label{eq:c_to_h}

\end{lem}

Suppose $g$ is a real number in $C_{n-1} \cap \interval[{k_{n-1}, k_{n-1}^*})$. 

\[g \in C_{n-1} \cap \interval[{k_{n-1}, k_{n-1}^*})\]
\[g \in C \cap V_2 \cap V_3 \cap ... \cap V_{n-1} \cap \interval[{k_{n-1}, k_{n-1}^*})\]

Since $k_{q-1}^* < g < k_{n-1}^*$, there must be some $j \leq n-1$ such that $k_{j-1}^* \leq g < k_j^*$

\[g \in C \cap V_2 \cap V_3 \cap ... \cap V_j \cap \interval[{k_{n-1}, k_{n-1}^*})\]
\[g \in C_j^* \cap \interval[{k_{j-1}^*, k_j^*}) \cap \interval[{k_{n-1}, k_{n-1}^*})\]
\[g \in C_H^* \cap \interval[{k_{n-1}, k_{n-1}^*})\]

As this is true for any real number in $C_{n-1} \cap \interval[{k_{n-1}, k_{n-1}^*})$,

\[C_{n-1} \cap \interval[{k_{n-1}, k_{n-1}^*}) \subseteq C_H^* \cap \interval[{k_{n-1}, k_{n-1}^*})\]

\begin{lem}

For any integer $j$, if $q \leq j \leq n-1$, $|\interval[{k_{j-1}^*, k_j^*}) \cap V_1^* \cap C_j^*| \leq |\interval[{k_{j-1}, k_j}) \cap V_1 \cap C_j|$
\label{eq:zero}
\end{lem}

For any integer $j$, $q \leq j \leq n-1$, we have from the split decision algorithm applied to the division of $C_j$

\[|\interval[{k_{j-1}, k_j}) \cap V_1 \cap C_j| = \frac{w_{j+1}}{w_1^j}|\interval[{k_j, \infty}) \cap C_j \cap V_{j+1}|\]
\[|\interval[{k_{j-1}^*, k_j^*}) \cap V_1^* \cap C_j^*| = \frac{w_{j+1}}{w_1^j}|\interval[{k_j^*, \infty}) \cap C_j^* \cap V_{j+1}|\]

Where $w_1^j$ is the workload assigned to $P_1$ in the division of $C_j$.

\[|\interval[{k_{j-1}, k_j}) \cap V_1 \cap C_j| = \frac{w_{j+1}}{w_1^j}|\interval[{k_j, \infty}) \cap V_2 \cap V_3 \cap ... \cap V_j\cap V_{j+1}|\]
\[|\interval[{k_{_j-1}^*, k_j^*}) \cap V_1^* \cap C_j^*| = \frac{w_{j+1}}{w_1^j}|\interval[{k_j^*, \infty}) \cap V_2 \cap V_3 \cap ... \cap V_j \cap V_{j+1}|\]

As $k_j^* \geq k_j$, $\interval[{k_j^*, \infty}) \subseteq \interval[{k_j, \infty})$, and we have

\[  |\interval[{k_{j-1}^*, k_j^*}) \cap V_1^* \cap C_j^*| \leq |\interval[{k_{j-1}, k_j}) \cap V_1 \cap C_j| \]

\begin{lem}

$|V_1 \cap \bar{V_1^*} \cap C_H^* \cap \interval[{k_{q-1}, k_{n-1}})| \geq |V_1^* \cap C_H^* \cap \interval[{k_{n-1}, k_{n-1}^*})|$
\label{eq:two}
\end{lem}

Using lemma \ref{eq:h_to_h}, we have

\[|\interval[{k_{q-1}, k_{n-1}}) \cap V_1 \cap C_H^* |\geq |\interval[{k_{q-1}, k_{n-1}}) \cap V_1 \cap C_H|\]
\[|\interval[{k_{q-1}, k_{n-1}}) \cap V_1 \cap C_H^*| \geq |V_1 \cap C_q \cap \interval[{k_{q-1}, k_q})| + |V_1 \cap C_{q+1} \cap \interval[{k_q, k_{q + 1}})| +... +  |V_1 \cap C_{n-1} \cap \interval[{k_{n-2}, k_{n - 1}})|\]

Using lemma \ref{eq:zero} on each of those terms,

\[|\interval[{k_{q-1}, k_{n-1}}) \cap V_1 \cap C_H^*| \geq |V_1^* \cap C_q^* \cap \interval[{k_{q-1}^*, k_q^*})| + |V_1^* \cap C_{q+1}^* \cap \interval[{k_q^*, k_{q + 1}^*})| +... +  |V_1^* \cap C_{n-1}^* \cap \interval[{k_{n-2}^*, k_{n - 1}^*})|\]
\[|\interval[{k_{q-1}, k_{n-1}}) \cap V_1 \cap C_H^*| \geq |V_1^* \cap C_H^* \cap \interval[{k_{q-1}^*, k_{n-1}^*})|\]

Using the fact that $k_{q-1}^* < k_{q-1}$

\[|\interval[{k_{q-1}, k_{n-1}}) \cap V_1 \cap C_H^*| \geq |V_1^* \cap C_H^* \cap \interval[{k_{q-1}, k_{n-1}^*})|\]

Splitting the right side at $k_{n-1}$,

\[|\interval[{k_{q-1}, k_{n-1}}) \cap V_1 \cap C_H^*| \geq |V_1^* \cap C_H^* \cap \interval[{k_{q-1}, k_{n-1}})| + |V_1^* \cap C_H^* \cap \interval[{k_{n-1}, k_{n-1}^*})|\]
\[|\interval[{k_{q-1}, k_{n-1}}) \cap V_1 \cap C_H^*| - |V_1^* \cap C_H^* \cap \interval[{k_{q-1}, k_{n-1}})| \geq |V_1^* \cap C_H^* \cap \interval[{k_{n-1}, k_{n-1}^*})|\]

For any two segments $A$ and $B$, $|A \cap \bar{B}| \geq |A| - |B|$.

\[|V_1 \cap \bar{V_1^*} \cap C_H^* \cap \interval[{k_{q-1}, k_{n-1}})| \geq |V_1^* \cap C_H^* \cap \interval[{k_{n-1}, k_{n-1}^*})|\]

\begin{lem}

If $k_{n-1}^* \geq k_{n-1}$, $|V_1 \cap D_1| \leq |V_1 \cap D_1^*|$

\end{lem}

Now we can put it all together.

\[|V_1 \cap D_1| = |\interval[{k_{n-1}, \infty}) \cap C_{n-1} \cap V_n \cap V_1|\]
\[|V_1 \cap D_1| = |\interval[{k_{n-1}, k_{n-1}^*}) \cap C_{n-1} \cap V_n \cap V_1| + |\interval[{k_{n-1}^*, \infty}) \cap C_{n-1} \cap V_n \cap V_1|\]
\[|V_1 \cap D_1| \leq |\interval[{k_{n-1}, k_{n-1}^*}) \cap C_{n-1} \cap V_1| + |\interval[{k_{n-1}^*, \infty}) \cap C_{n-1} \cap V_n \cap V_1|\]

Using lemma \ref{eq:c_to_h},

\[|V_1 \cap D_1| \leq |\interval[{k_{n-1}, k_{n-1}^*}) \cap C_H^* \cap V_1| + |\interval[{k_{n-1}^*, \infty}) \cap C_{n-1} \cap V_n \cap V_1|\]
\[|V_1 \cap D_1| \leq |\interval[{k_{n-1}, k_{n-1}^*}) \cap C_H^* \cap V_1 \cap \bar{V_1^*}| + |\interval[{k_{n-1}, k_{n-1}^*}) \cap C_H^* \cap V_1 \cap V_1^*| + |\interval[{k_{n-1}^*, \infty}) \cap C_{n-1} \cap V_n \cap V_1|\]
\[|V_1 \cap D_1| \leq |\interval[{k_{n-1}, k_{n-1}^*}) \cap C_H^* \cap V_1 \cap \bar{V_1^*}| + |\interval[{k_{n-1}, k_{n-1}^*}) \cap C_H^* \cap V_1^*| + |\interval[{k_{n-1}^*, \infty}) \cap C_{n-1} \cap V_n \cap V_1|\]

Using lemma \ref{eq:two}, 

\[|V_1 \cap D_1| \leq |\interval[{k_{n-1}, k_{n-1}^*}) \cap C_H^* \cap V_1 \cap \bar{V_1^*}| + |\interval[{k_{q-1}, k_{n-1}}) \cap C_H^* \cap V_1 \cap \bar{V_1^*}| + |\interval[{k_{n-1}^*, \infty}) \cap C_{n-1} \cap V_n \cap V_1|\]
\[|V_1 \cap D_1| \leq |\interval[{k_{q-1}, k_{n-1}^*}) \cap C_H^* \cap V_1 \cap \bar{V_1^*}| + |\interval[{k_{n-1}^*, \infty}) \cap C_{n-1} \cap V_n \cap V_1|\]

$C_{n-1} \cap V_n \cap V_1$ is the intersection of $\interval[{k_{n-1}, \infty})$ with all $n$ valuation sets. $ C_{n-1}^* \cap V_n \cap V_1$ is the same but to the right of $k_{n-1}^*$. As $k_{n-1}^* \geq k_{n-1}$, $\interval[{k_{n-1}^*, \infty}) \cap C_{n-1} \cap V_n \cap V_1 = \interval[{k_{n-1}^*, \infty}) \cap C_{n-1}^* \cap V_n \cap V_1$.

\[|V_1 \cap D_1| \leq |\interval[{k_{q-1}, k_{n-1}^*}) \cap C_H^* \cap V_1 \cap \bar{V_1^*}| + |\interval[{k_{n-1}^*, \infty}) \cap C_{n-1}^* \cap V_n \cap V_1|\]
\[|V_1 \cap D_1| \leq |V_1 \cap ((\interval[{k_{q-1}, k_{n-1}^*}) \cap C_H^* \cap \bar{V_1^*}) \cup (\interval[{k_{n-1}^*, \infty}) \cap C_{n-1}^* \cap V_n))|\]
\[|V_1 \cap D_1| \leq |V_1 \cap D_1^*|\]

With both the $k_{n-1}^* < k_{n-1}$ and $k_{n-1}^* \geq k_{n-1}$ cases covered, we have that no matter the $V_1^*$, $P_1$ can never benefit from deception.

\begin{lem}

For any input to the split decision algorithm $C, n, V, w$ and any valuation set $V_2^*$, if $V^*$ is the same as $V$ expect that $V_2$ is replaced by $V_2^*$, if $\{D_1, D_2, ... ,D_n\} = f(C, n, V, w)$, and $\{D_1^*, D_2^*, ..., D_n^*\} = f(C, n, V^*, w)$, $|V_2 \cap D_2| \leq |V_2 \cap D_2^*|$

\end{lem}

The same argument as for $P_1$ applies for $P_2$, with left and right swapped.

\begin{thm}

The split rulership algorithm is strategyproof

\end{thm}

As we have solved the cases of $P_1$ and $P_2$, all that remains is the other $n-2$ players. Suppose that $P_i$ submits some $V_i^*$, $i > 2$, with $V^*$ the same as $V$ except $V_i$ is replaced by $V_i^*$. In the split decision algorithm, only $V_1$ and $V_2$ are used in the choice of $k$ and creation of $C_L$ and $C_R$, and so when the algorithm is run with $V^*$ as input the same $C_L$ and $C_R$ are produced. Let $D_i^L$ and $D_i^R$ be the chores assigned to $P_i$ in the divisions of $C_L$ and $C_R$ when the input is $V$, and $D_i^{L*}$ and $D_i^{R*}$ be the same but with $V^*$ as input. We may proceed by induction on the number of players. Taking as given that the split decision algorithm is strategyproof when dividing among $n-1$ players, we have that

\[|V_i \cap D_i^L| \leq |V_i \cap D_i^{L*}| \text{ and } |V_i \cap D_i^R| \leq |V_i \cap D_i^{R*}|\]
\[|V_i \cap D_i| \leq |V_i \cap D_i^*|\]

\section{Discussion}

A chore division algorithm is \textit{envy-free} if, when run with $w = \{\frac{1}{n}, \frac{1}{n},..., \frac{1}{n}\}$ each player believes that, not only is their assigned chores at most $\frac{1}{n}$ of the total chores, but they have the least costly assigned chores among all players. That is, for any two players $P_i$ and $P_j$, $|V_i \cap D_i| \leq |V_i \cap D_j|$. The split rulership algorithm is not envy-free. In general, envy-freeness is harder to achieve for chore division than for cake division, and I suspect that envy-freeness and strategyproofness are incompatible for chore division.

A chore division algorithm is \textit{Pareto efficient} if there is no alternate way the chores could be divided in which at least one player is more satisfied and no players are less satisfied. That is, if the algorithm produces the division $\{D_1, D_2, ... D_n\}$, for any other partition of $C$ $\{D_1', D_2', ... D_n'\}$, if there is some player $P_i$ such that $|V_i \cap D_i| > |V_i \cap D_i'|$, there must also be some player $P_j$ such that $|V_i \cap D_i| < |V_i \cap D_i'|$. The split rulership algorithm is Pareto efficient. The only chores in the $V_i \cap D_i$ are those chores that all players have marked, so the total cost paid by all players is $|V_1 \cap V_2 \cap ... \cap V_n|$. Any other partition of the chores must have a total cost at least this great. However, if at least one player paid less cost and no players paid more cost, the total cost paid would be lower.  

Piecewise uniform valuation functions are the most restrictive class of valuation functions for which chore division as a problem makes sense. Other classes of valuation functions that have been proposed include piecewise constant (in which each interval has a constant marginal cost, but those marginal costs need not be equal) and piecewise linear (in which each interval has a marginal cost that is a linear function). A very simple extension might have players' marginal valuations restricted to a finite set of values. No deterministic strategyproof algorithms have been found for these less restrictive class of valuation functions for either cake cutting or chore division.

The split rulership algorithm is incompatible with the Robertson-Webb query model of algorithm complexity (\cite{robertson1998cake}). The placement of $k$ is dependent on both $P_1$ and $P_2$'s valuations; no finite number of cut and evaluation queries is guaranteed to find it. \cite{kurokawa2013cut} found that no strategyproof, envy-free algorithm for cake cutting with piecewise uniform valuation functions using a bounded number of queries exists; it is possible a similar result is true for chore division.

\bibliographystyle{unsrtnat}
\bibliography{strategyproof_f2}  





\end{document}